\documentclass[
    ,final            % use final for the camera ready runs
  ]
  {aipproc}

\layoutstyle{8x11double}

\begin{document}

\title{Birth and Evolution of Isolated Radio Pulsars}


\author{Claude-Andr\'e Faucher-Gigu\`ere}{
  address={Department of Astronomy, Harvard University, 60 Garden St, MS-10, Cambridge, MA, 02138, USA; cgiguere@cfa.harvard.edu.}
}

\author{Victoria M. Kaspi}{
  address={Department of Physics, McGill University, 3600 University St, Montr\'eal, QC, H3A-2T8, Canada.}
}

\begin{abstract}
We investigate the birth and evolution of isolated radio pulsars using a population synthesis method, modeling the birth properties of the pulsars, their time evolution, and their detection in the Parkes and Swinburne Multibeam (MB) surveys. Together, the Parkes and Swinburne MB surveys 
\citep[][]{2001MNRAS.328...17M, 2001MNRAS.326..358E} have detected nearly 2/3 of the known pulsars and provide a remarkably homogeneous sample to compare with simulations. New proper motion measurements \citep[][]{2002ApJ...571..906B, 2003AJ....126.3090B} and an improved model of the distribution of free electrons in the interstellar medium, NE2001 \citep[][]{NE2001}, also make revisiting these issues particularly worthwhile. We present a simple population model that reproduces the actual observations well, and consider others that fail. We conclude that: pulsars are born in the spiral arms, with the birthrate of $2.8\pm0.5$ pulsars/century peaking at a distance $\sim3$ kpc from the Galactic centre, and with mean initial speed of $380^{+40}_{-60}$ km s$^{-1}$; the birth spin period distribution extends to several hundred milliseconds, with no evidence of multimodality, implying that characteristic ages overestimate the true ages of the pulsars by a median factor >2 for true ages <30,000 yr; models in which the radio luminosities of the pulsars are random generically fail to reproduce the observed $P-\dot{P}$ diagram, suggesting a relation between intrinsic radio luminosity and $(P, \dot{P})$; radio luminosities $L \propto \sqrt{\dot{E}}$ provides a good match to the observed $P-\dot{P}$ diagram; for this favored radio luminosity model, we find no evidence for significant magnetic field decay over the lifetime of the pulsars as radio sources ($\sim100$ Myr).
\end{abstract}

%%%%%%%%%%%%%%%%%%%%%%%%%%%%%%%%%%%%%%%%%%%%%%%%%%%%%%%%%%%%%%%%%%%
%%
%% The below \maketitle command inserts the actual front matter data.
%% It has to follow the above declarations.
%%
%%%%%%%%%%%%%%%%%%%%%%%%%%%

\maketitle

%%%%%%%%%%%%%%%%%%%%%%%%%%%%%%%%%%%%%%%%%%%%
%% MAINMATTER
%%
%%%%%%%%%%%%%%%%%%%%%%%%%%%%%%%%%%%%%%%%%%%%%%%%%%%%%%%%%%%%%%%%%%%%%%%%%%%%
%% Headings:
%%
%% The aipproc supports three heading levels, i.e., \section,
%%	\subsection, and \subsubsection.
%%%%%%%%%%%%%%%%%%%%%%%%%%%%%%%%%%%%%%%%%%%%%%%%%%%%%%%%%%%%%%%%%%%%%%%%%%%%
%% Cross-references:
%%
%% Page numbers (\pageref) and headings can NOT be referenced in the class,
%% since before being produced, no page numbers are determined.
%%
%% Tables, figures, and equeations can be referenced by using the LaTex
%% 	commands \label and \ref. For references to equation numbers, \eqref
%%	can be used, which will print "(1)" (while \ref will result in "1").
%%
%%%%%%%%%%%%%%%%%%%%%%%%%%%%%%%%%%%%%%%%%%%%%%%%%%%%%%%%%%%%%%%%%%%%%%%%%%%%
%% Lists: 
%%
%% Standard "itemize", "enumerate", etc. list environments are supported.
%%%%%%%%%%%%%%%%%%%%%%%%%%%%%%%%%%%%%%%%%%%%%%%%%%%%%%%%%%%%%%%%%%%%%%%%%%%%
%% Urls:
%%
%% \url{} command is provided for documenting URLs.
%%%%%%%%%%%%%%%%%%%%%%%%%%%%%%%%%%%%%%%%%%%%

\section{INTRODUCTION}
From soon after the discovery of the pulsars \citep[][]{1968Natur.217..709H}, their Galactic population has been the focus of numerous studies \citep[][]{1970ApJ...160..979G, 1971IAUS...46..165L, 1977ApJ...215..885T, 1977MNRAS.179..635D, 1985MNRAS.213..613L, 1987A&A...171..152S, 1989ApJ...345..931E, 1992A&A...254..198B, 1993MNRAS.263..403L, 1997A&A...322..127H, 1997MNRAS.289..592L, 2002ApJ...568..289A, 2002ApJ...565..482G, 2004ApJ...604..775G}.
Nevertheless, in spite of much progress, many outstanding questions remain.
What are the birth positions, velocities, spin periods, and magnetic fields of the pulsars?
How do these evolve in time and how are they related to the radio luminosities of the pulsars?
One approach to answering these questions is to make use of the statistical power of the growing pulsar catalogue to study the pulsar population as a whole.\\ \\ 
Many recent advances in pulsar astronomy make it particularly worthwhile to revisit the above questions through population synthesis.
The recently completed Parkes and Swinburne Multibeam (PMB and SMB; \cite{2001MNRAS.328...17M, 2001MNRAS.326..358E}) surveys have detected nearly 2/3 of the known pulsars and provide a large and remarkably homogeneous observed sample to compare with simulations.
New proper motion measurements \citep[][]{2002ApJ...571..906B, 2003AJ....126.3090B} and an improved model of the interstellar medium \citep[][]{NE2001} also provide valuable new information.\\ \\
In this work, the details of which have been reported by Faucher-Gigu\`ere \& Kaspi (2006) \cite{2006ApJ...643..332F}, we investigate the birth properties of Galactic isolated radio pulsars and their time evolution.
To do so, we generate an ensemble of mock galaxies populated by pulsars with prescribed birth properties (spatial locations, velocities, spin periods, radio luminosities, magnetic fields) and evolve the pulsars in time using physical models.
We model the selection function of the PMB and SMB surveys using a modified version of the radiometer equation \citep[][]{1985ApJ...294L..25D} and apply it to our mock galaxies.
We compare the observed histograms of Galactic longitudes, latitudes, dispersion measures, 1.4 GHz radio fluxes, pulse periods, magnetic fields, as well as the observed $P-\dot{P}$ diagrams, to judge how well each population model reproduces the actual observations.

\begin{figure*}
  \includegraphics[height=.25\textheight]{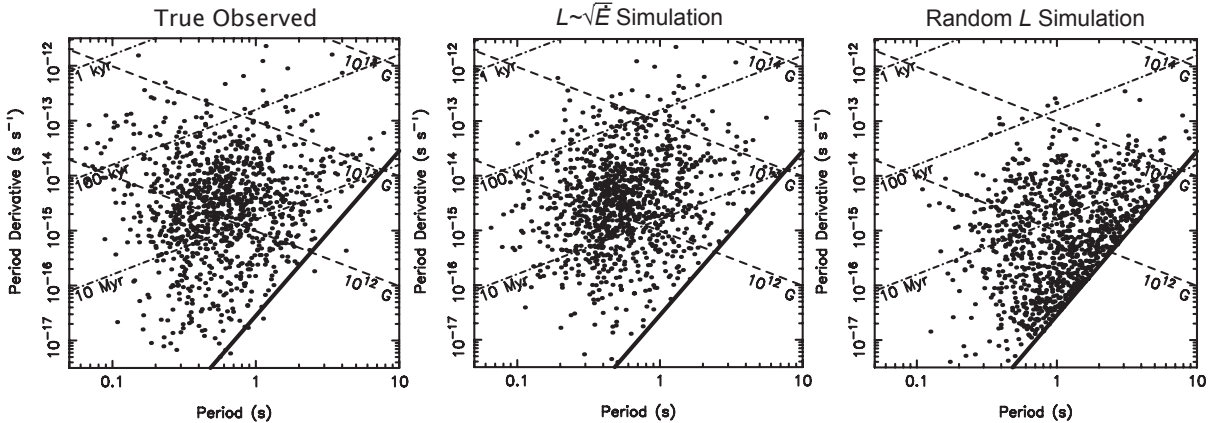}
  \caption{Comparison of the $P-\dot{P}$ diagram for the pulsars detected in the Parkes and Swinburne Multibeam surveys (left) with the corresponding diagrams in our best simulation with $L\propto \sqrt{\dot{E}}$ (middle) and with a simulation in which the radio luminosities of the pulsars are uncorrelated with their other characteristics (``random''; right). 
While the simulation with $L\propto \sqrt{\dot{E}}$ reproduces the observed diagram remarkably well, the model with random radio luminosities produces a pile-up of pulsars near the death line that is in clear disagreement with the observations.}
\label{ppdot diagrams}
\end{figure*}

\section{RESULTS}
\begin{figure*}
  \includegraphics[height=.46\textheight,angle=-90]{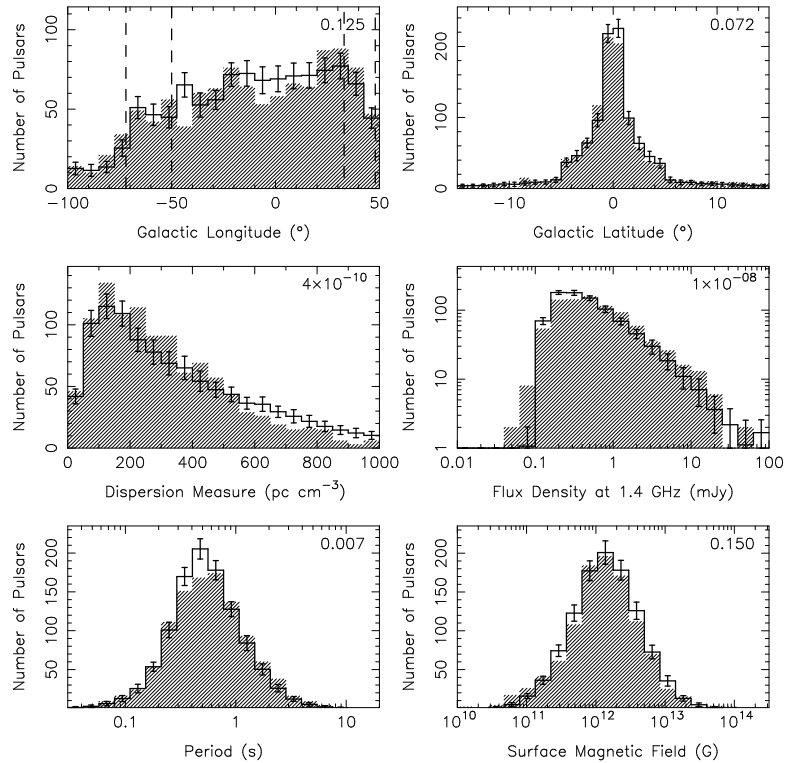}
  \caption{Distributions of observed pulsar Galactic longitudes and latitudes, dispersion measures, flux densities at 1.4 GHz, pulse periods, and surface magnetic fields for our best model with $L\propto \sqrt{\dot{E}}$ (solid lines) compared to the real distributions (hatched histograms).}
\label{distributions}
\end{figure*}

In Figure \ref{ppdot diagrams}, we compare the $P-\dot{P}$ diagram for the pulsars detected in the PMB and SMB surveys  with the corresponding diagrams in our best simulation with radio luminosities $L\propto \sqrt{\dot{E}}$ and with a simulation in which the radio luminosities of the pulsars are uncorrelated with their other characteristics (``random'').
In Figure \ref{distributions}, we compare the histograms of Galactic longitudes, latitudes, dispersion measures, 1.4 GHz radio fluxes, pulse period, and magnetic fields obtained in our best simulation with the actual observations.
In what follows, we summarize the key results obtained from our preferred model and its comparison with various alternatives.\\ \\
We find that pulsars are born at a rate of $2.8 \pm 0.5$ per century in the Galaxy, with the rate peaking a distance of $\sim3$ kpc from the Galactic centre (in agreement with \cite{lor03, 2004A&A...422..545Y}), and a mean velocity of $380^{+40}_{-60}$ \mbox{km s$^{-1}$}.
In particular, models in which the pulsar birthrate peaks at the Galactic centre or is uniform throughout the Galactic disk, as is frequently assumed (e.g., \cite{1993MNRAS.263..403L, 2002ApJ...568..289A}), fail to reproduce the observed distribution of Galactic longitudes.
We further find evidence that the pulsar birthrate traces the Galactic spiral arms, as expected if pulsars are formed in core collapse supernovae marking the death of short-lived massive stars.\\ \\
The birth spin period distribution extends to several hundred milliseconds, with no evidence of multimodality.
As a consequence, the assumption that $P/P_{0}\ll 1$ is violated for many young pulsars and the characteristic age $t_{char}\equiv P/2\dot{P}$ is an overestimate of the true age of a pulsar by a median factor $>2$ for true ages $<30,000$ yr.
This is consistent, for example, with PSR J1811$-$1925 having a characteristic age $\sim 12 \times$ the age inferred from its association with the supernova remnant G$11.2-0.3$ (AD 386; \cite{1999ApJ...523L..69T, 2001ApJ...560..371K}).
\\ \\
Models in which the pulsar radio luminosities, $L$, are randomly assigned to pulsars generically predict too many pulsars detected away from the Galactic plane and a clear pile-up on the death line in the $P-\dot{P}$ diagram that is not observed (Fig. \ref{ppdot diagrams}).
This suggests a relation between a pulsar's intrinsic radio luminosity and its $(P,~\dot{P})$ which favors the detection of young pulsars, as was often assumed in previous analyses (e.g., \cite{1993MNRAS.263..403L}).
We find that $L\propto \sqrt{\dot{E}}$ naturally solves both issues by uniformly dimming the pulsars as they age and approach the death line.\footnote{In this model, the radio luminosity contours are parallel to the death line.}\\ \\
Finally, in contrast with previous studies following a similar methodology (e.g., \cite{2002ApJ...565..482G, 2004ApJ...604..775G}), we do not find evidence for significant magnetic field decay over the lifetime of pulsars as radio sources ($\sim$ 100 Myr).
Indeed, we have throughout our study \emph{assumed} that the magnetic fields of our mock pulsars were constant in time and were able to obtain good agreement for our best model in which $L\propto \sqrt{\dot{E}}$ between the simulations and the real pulsars (Fig. \ref{ppdot diagrams}, \ref{distributions}).
We caution, however, that this does not constitute a proof of that pulsar magnetic fields are constant.
Rather, much of the success of our favored model lies in the chosen radio luminosity dependence on $P$ and $\dot{P}$ and it is possible that another choice would produce similar agreement.
In the absence of an independently determined luminosity model, we must regardless conclude that pulsar population synthesis studies currently do not \emph{require} any decay of pulsar magnetic fields.

\begin{theacknowledgments}
CAFG acknowledges support from NSERC.
VMK is a Canada Research Chair and the Lorne Trottier Chair in Astrophysics and Cosmology.
%This work has made extensive use of a computer cluster purchased with funds of the Canada Foundation for Innovation.
Further support was provided by NSERC, CFI, FQRNT, and CIFAR.
%the Fonds qu\'eb\'ecois de la recherche sur la nature et les technologies, and the Canadian Institute for Advanced Research.

\end{theacknowledgments}

%%%%%%%%%%%%%%%%%%%%%%%%%%%%%%%%%%%%%%%%%%%%%%%%
%% The bibliography can be prepared using the BibTeX program or
%% manually.
%%
%% The code below assumes that BibTeX is used. Compliant BibTex styles
%% are aipproc (for use with natbib) and aipprocl (if natbib is missing
%% at the site).
%%
%% Please run "bibtex \jobname" to obtain the bibliography and 
%% then re-run LaTeX twice to fix the references!
%%
%% When referring to citations in the text, in quare brackets [] show
%% the number in order of appearance. References in the References
%% section are listed in the same numerical order.
%%%%%%%%%%%%%%%%%%%%%%%%%%%%%%%%%%%%%%%%%%%%%%%%%

\bibliographystyle{aipproc}   % if natbib is available
%\bibliographystyle{aipprocl} % if natbib is missing

%%%%%%%%%%%%%%%%%%%%%%%%%%%%%%%%%%%%%%%%%%%
%% You probably want to use your own bibtex database here
%%%%%%%%%%%%%%%%%%%%%%%%%%%%%%%%%%%%%%%%%%%

\bibliography{references}

%%%%%%%%%%%%%%%%%%%%%%%%%%%%%%%%%%%%%%%%%%%%%%%%%
%% If the bibliography is
%% produced without BibTeX, comment out the above lines, use
%% \begin{thebibliography}{widest-label} environment to hold 
%% the list of references and 
%% \bibitem{label} command to start a bibliographical entry having
%% the "label" for use in \cite commands.
%%
%% For your convenience a manually coded example is appended
%% after the \end{document}
%%%%%%%%%%%%%%%%%%%%%%%%%%%%%%%%%%%%%%%%%%%%%%%%

\end{document}